\documentstyle[12pt]{article}
\def\barr{\begin{array}}
\def\earr{\end{array}}
\def\beq{\begin{equation}}
\def\eeq{\end{equation}}
\def\bea{\begin{eqnarray}}
\def\eea{\end{eqnarray}}
\def\bmath{\begin{displaymath}}
\def\emath{\end{displaymath}}
\def\bq{\begin{quote}}
\def\eq{\end{quote}}
\def\slash#1{\setbox0=\hbox{$#1$}#1\hskip-\wd0\hbox to\wd0{\hss\sl/\/\hss}}
\begin{document}

\begin{flushright}
MZ-TH/93-18 \\
July 1993
\end{flushright}

\begin{center}
{\bf{\Large ASTROPHYSICAL AND TERRESTRIAL }} \\[0.3cm]
{\bf{\Large CONSTRAINTS }} \\[0.3cm]
{\bf{\Large ON SINGLET MAJORON MODELS}}\\[1.cm]
{\large Apostolos~Pilaftsis}$^{\displaystyle \, \, (a) ,}$
\footnote[1]{Address after 1, Oct.~1993, Rutherford Appleton Laboratory,
Chilton, Didcot, Oxon,\\
{\em ENGLAND}. E-mail address: pilaftsis@vipmza.physik.uni-mainz.de}\\
$^{(a)}$ Institut f\"ur Physik, Johannes-Gutenberg Universit\"at,
Staudinger Weg 7,\\
55099 Mainz, {\em FRG}
\end{center}
\bigskip
\bigskip
\centerline {\bf ABSTRACT}

The general Lagrangian containing the couplings of the Higgs scalars to
Majorana neutrinos is presented in the context of singlet Majoron models
with intergenerational mixings. The analytical expressions for the coupling
of the Majoron field to fermions are derived within these models.
Astrophysical considerations imply severe restrictions on the
parameters of the model if the singlet Majoron model with three
generations is assumed to be embedded in grand unified theories.
Bounds that originate from analyzing possible charged lepton-violating decays
in terrestrial experiments are also discussed. In particular, we find that
experimental searches for muon decays by Majoron emission cannot generally
be precluded by astrophysical requirements.

\newpage

Astrophysical considerations play an important role in constraining the
strength of the coupling of Nambu-Goldstone bosons to the matter~\cite{GR}.
Among the various types of such extraordinary light particles (e.g.~axions,
familons, etc.)~\cite{GNY},
which are acossiated with the spontaneous breakdown of some global symmetry,
Majoron $J^0$ is a massless pseudoscalar boson arising from the breaking
of the baryon$-$lepton ($B-L$) symmetry~\cite{CMP}-\cite{JV}.
In such scenarios, apart from the
Standard Model ($SM$) Higgs doublet $\Phi$, an
$SU(2)\otimes U(1)$ singlet $\Sigma$ is present which gives rise to
$\Delta L=2$ Majorana mass terms $m_M$ when $\Sigma$ couples to right-handed
neutrinos~\cite{CMP}.
In general, models with right-handed neutrinos can naturally
account for possible lepton-number violating decays of $Z^0$ and $H^0$
particles~\cite{APH}-\cite{BP}
induced by Majorana neutrinos at the first electroweak
loop order. The non-decoupling physics that the heavy Majorana neutrinos
can introduce~\cite{APH,KPS,BP}
lead to combined constraints both on neutrino masses
and lepton-violating mixings. Similar non-decoupling effects occur when one
considers Majoron couplings to fermions. On the other hand, Majoron models
can naturally be embedded in grand unified theories ($GUT$) like the
$SO(10)$ model~\cite{GUT}. In such models the $B-L$ scale is determined
by the vacuum
expectation value ($VEV$) of the singlet scalar, i.e.~$<\Sigma > = w/\sqrt{2}$
and the neutrino Dirac mass matrix $m_D$ may be related to the $u$-quark mass
matrix $M_U$ by $m_D = M_U/k$, where $k = {\cal O}(1)$ represents the
running of the Yukawa couplings between the $GUT$ and the low-energy
scale~\cite{BKL}.
There are also scenarios where $m_D$ can be proportional to charged lepton
mass matrix $M_l$~\cite{GUT,BKL}. Since the $B-L$ scale strongly depends on
the mechanism that the $SO(10)$ gauge group breaks down to
$U(1)_{em}$~~\cite{DC}, we will treat $w$ as a free parameter of the theory
that should be constrained by our forthcoming considerations.
The light neutrino masses $m_{\nu_i}$ can generally be estimated by
\beq
m_{\nu_i}\ \ \simeq \ \ -\ m_D\ \frac{1}{m_M}\ m_D^T.
\eeq 
The mass hierarchical pattern mentioned above should also hold for the heaviest
family which implies that the biggest eigenvalue of $m_D$ will be in the
range between 10 and 100~GeV for $100\leq m_t\leq 180$~GeV~\cite{BKL}.
If astrophysical constraints are imposed on the $J^0-e-e$,
$J^0-u-u$ and $J^0-d-d$ couplings, one derives the bound on
\beq
\tan \beta\ \ =\ \ \frac{<\Phi >}{<\Sigma >}\ \ =\ \ \frac{v}{w}\ \ \leq\ \
10^{-2},
\eeq 
for models without interfamily mixings. The situation becomes more
involved  if mixings between families are considered. This realization
will be illustrated by the present work.
We will finally discuss the bounds from low-energy
experiments~\cite{TER1,TER2}
on the off-diagonal coupling of the Majoron to two different charged leptons.

First, let us briefly describe the low-energy structure of the singlet
Majoron model.
The scalar potential of this model which should be non-trivial under the $
U(1)_Y$ group is given by~\cite{HLP,JR}
\bea
-\ {\cal L}_V\ &=&\ \mu^2_\Phi (\Phi^\dagger \Phi)\ +\ \mu^2_\Sigma
(\Sigma^\dagger \Sigma)\ +\ \frac{\lambda_1}{2} (\Phi^\dagger \Phi)^2 \ +\
\frac{\lambda_2}{2}(\Sigma^\dagger \Sigma)^2 \nonumber\\
&&+\ \delta (\Phi^\dagger \Phi)(\Sigma^\dagger \Sigma).
\eea 
If all the stability conditions in this model are satisfied
(i.e.~$\lambda_1,\ \lambda_2 > 0$ and $\lambda_1\lambda_2 >
\delta$)~\cite{HLP}, the above potential can always be minimized
by the following Higgs-field configurations:
\beq
\Phi\ =\ \left ( \barr{c} G^+ \\ \displaystyle{\frac{v}{\sqrt{2}} +
\frac{\phi^0 + iG^0}{\sqrt{2}}} \earr \right) \qquad \mbox{and} \qquad
\Sigma\ =\ \frac{w}{\sqrt{2}}\ +\ \frac{\sigma^0 + iJ^0}{\sqrt{2}} .
\eeq
After the spontaneous breakdown of the $SU(2)_L\otimes U(1)_Y$ gauge group and
diagonalizing the Higgs mass matrix, one obtains two $CP$-even Higgs fields
(denoted by $H^0$ and $S^0$) and one massless $CP$-odd scalar, the Majoron
$J^0$, while the would-be Goldstone bosons $G^+$,
$G^0$ give simply mass to $W^+$, $Z^0$ bosons, respectively. The weak
eigenstates $\phi^0$ and $\sigma^0$ are related to the corresponding
physical mass eigenstates through
\beq
\left( \barr{c} \phi^0 \\ \sigma^0 \earr \right)\ =\
\left( \barr{cc} \cos \theta & - \sin \theta \\
                 \sin \theta & \cos\theta \earr \right)\ \left(
\barr{c} H^0 \\ S^0 \earr \right) \ ,
\eeq 
where
\beq
\tan 2\theta\ \ = \ \ \frac{2\delta \tan\beta}{\lambda_2 - \lambda_1
\tan^2\beta}\ .
\eeq 
The Yukawa sector containing all the relevant Higgs couplings to neutrinos
reads
\bea
- {\cal L}_Y^{Higgs} \ &=&\ \bar{\nu}^0_{L_i} m_{D_{ij}} \nu^0_{R_j}
\frac{\phi^0}{v}\ +\ \bar{\nu}^0_{R_i} m_{D_{ij}}^\dagger \nu^0_{L_j}
\frac{\phi^0}{v}\nonumber\\
&&\ +\ \frac{1}{2}\ \bar{\nu}^{0C}_{R_i} m_{M_{ij}} \nu^0_{R_j}
\ \frac{\sigma^0+iJ^0}{w}\ +\
\frac{1}{2}\ \bar{\nu}^{0}_{R_i} m_{M_{ij}}^\dagger \nu^{0C}_{R_j}
\ \frac{\sigma^0-iJ^0}{w}\ .
\eea 
In Eq.~(7) we have assumed the absence of Higgs triplets~\cite{GRTR},
which seem now to be ruled out by the $LEP$ data on the invisible $Z^0$ width.
The interactions
of $J^0$, $H^0$ and $S^0$ with the $2n_G$ Majorana neutrinos $n_i$ -- $n_G$
denotes the number of generations -- are
generally described by the following Lagrangians:
\bea
{\cal L}^J_{int}\ &=&\ \frac{i g_W t_\beta}{4M_W}\ J^0\
\bar{n}_i \left[ \gamma_5 (m_{n_i}+m_{n_j})\left( \frac{1}{2}\delta_{ij}-
\mbox{Re}C_{ij} \right)\ +\ i(m_{n_i}-m_{n_j})\mbox{Im}C_{ij} \right] n_j\ ,
\\[0.5cm]
{\cal L}^H_{int}\ &=& -\ \frac{g_W}{4M_W}(c_\theta-s_\theta t_\beta)\
H^0\ \bar{n}_i \Bigg[ (m_{n_i}+m_{n_j})\left( \mbox{Re}C_{ij}
+\frac{t_\beta s_\theta \delta_{ij}}{2(c_\theta-s_\theta t_\beta)} \right)
\nonumber\\
&&+\ i\gamma_5 (m_{n_j}-m_{n_i})\mbox{Im}C_{ij} \Bigg] n_j\ ,\\[0.5cm]
{\cal L}^S_{int}\ &=&\ \frac{g_W}{4M_W}(s_\theta+c_\theta t_\beta)\
S^0\ \bar{n}_i \Bigg[ (m_{n_i}+m_{n_j})\left( \mbox{Re}C_{ij}
-\frac{t_\beta \delta_{ij}}{2(t_\theta+t_\beta)} \right)\nonumber\\
&&+\ i\gamma_5 (m_{n_j}-m_{n_i})\mbox{Im}C_{ij} \Bigg] n_j\ ,
\eea 
where we have used the abbreviations $s_x=\sin x$, $c_x=\cos x$, $t_x=\tan x$
and defined
\beq
C_{ij}\ =\ \sum\limits_{k=1}^{n_G}\ U^\nu_{ki}U^{\nu\ast}_{kj}\ .
\eeq 
The $2n_G\times 2n_G$ unitary matrix $U^\nu$ is responsible for the
diagonalization of the $2n_G\times 2n_G$ neutrino mass matrix $M^\nu$ which
is of the "see-saw" form~\cite{YGRS}
\beq
M^\nu\ \ =\ \ \left( \barr{cc} 0 & m_D \\
                              m_D^T & m_M \earr \right) .
\eeq 
The first $n_G$ eigenvalues of $M^\nu$ are identified with the ordinary light
neutrinos, $\nu_e$, $\nu_\mu$, etc., whereas the remaining $n_G$ Majorana
neutrino states are new particles provided in this model and should be
heavier than the $Z^0$ boson to escape detection at $LEP$ experiments. The
$2n_G$ neutral leptons $n_i$ are related to their weak eigenstates
$\nu^0_{L,R_i}$ and $\nu^{0C}_{L,R_i}$
through the following unitary transformations (assuming the convention of
summation for repeated indices):
\beq
\left( \barr{c} \nu_L^0 \\ \nu^{0C}_R \earr \right)_i \ =\
U^{\nu\ast}_{ij}\ n_{L_j}\ , \qquad
\left( \barr{c} \nu_L^{0C} \\ \nu^0_R \earr \right)_i \ =\
U^\nu_{ij}\ n_{R_j} \ .
\eeq 
It is now easy to see that in the limit of $\tan\beta \to 0$ and $\theta
\to 0$, the fields $S^0$ and $J^0$ decouple fully from matter and only one
Higgs field, $H^0$, couples to Majorana neutrinos.
This scenario has explicitly been
described in~\cite{ZPC}, where for our purposes we will repeate here the
interactions of the $W^+$ and $Z^0$ boson with the Majorana neutrinos. They
are given by the Lagrangians:
\bea
{\cal L}_{int}^{W^\mp}& = & -\frac{g_W}{2\sqrt{2}} W^{-\mu}\
{\bar{l}}_i \ B_{l_ij} {\gamma}_{\mu} (1-{\gamma}_5) \ n_j \quad + \quad h.c.\
,
\\[0.5cm]
{\cal L}_{int}^Z& = & -\frac{g_W}{4\cos\theta_W}  Z^{0\mu}\
\bar{n}_i \gamma_\mu [ i\mbox{Im}(C_{ij})\ -\ \gamma_5\mbox{Re}(C_{ij}) ]
n_j\ .
\eea 
Moreover, the couplings of the charged would-be Goldstone bosons $G^\mp$ are
written down
\beq
{\cal L}_{int}^{G^\mp}\  =  -\frac{g_W}{2\sqrt{2}M_W} G^-\
{\bar{l}}_i \ [m_{l_i}B_{l_ij} (1-{\gamma}_5)\ -\ B_{l_ij}(1+{\gamma}_5)
m_{n_j}] \ n_j\quad + \quad h.c.\ ,
\eeq 
where
\beq
B_{l_ij} = \sum\limits_{k=1}^{n_G} V^l_{l_ik} U^{\nu\ast}_{kj}\ .
\eeq 
In Eq.~(17) $V^l$ is a unitary matrix relevant for the bi-diagonalization
of the charged lepton mass matrix $M^l$.

At this point it is important to mention that the presence of Majorana
neutrino interactions in the Lagrangians~(8)--(10)
violates the $CP$ symmetry of the
model. In particular, the fact that $H^0$, $S^0$ and $J^0$ couple
simultaneously to $CP$-even (:$\bar{n}_i n_j$) and $CP$-odd (:$\bar{n}_i
i\gamma_5 n_j$) operators gives rise
to $CP$-violating transitions between states with
different $CP$-quantum numbers~\cite{CNP}. For example, one finds a non-zero
contribution when computing selfenergy graphs induced by Majorana
neutrinos between the $CP$-even Higgs fields $H^0$, $S^0$ and the $CP$-odd
states $Z^0$, $J^0$. All these transitions turn out to be proportional to the
$CP$-odd combinations $\mbox{Im}C^2_{ij}$ which change sign when a $CP$
conjugation is applied to the vacuum polarization terms. As a consequence, the
general Majoron coupling to charged leptons and quarks possesses a scalar and
pseudoscalar part. We will include these $CP$-odd effects in our theoretical
considerations, although they seem not to influence our numerical
predictions.

Armed with the Lagrangians (8)--(10), (14) and (15) it is now
straightforward to calculate the coupling $J^0-f_1-f_2$ given by the
Feynman graphs shown in Figs.~1(a)--1(c). The additional $CP$-violating
diagrams of the Majoron coupling to fermions are depicted in Figs.~2(d)
and~2(e). In our analytical calculations we have neglected terms proportional
to the small quantities $m_f^2/M^2_W$ for $f= e, u$ or $d$. The individual
amplitudes contributing to the $J^0-f_1-f_2$ coupling are given by
\bea
{\cal T}^{l_1l_2}_a\ &=&\ \Delta^S_{ij}\
\Bigg[-\frac{1}{2}(\lambda_i+\lambda_j)
(\delta_{ij} -C^\ast_{ij} )I_1(\lambda_i,\lambda_j)\ +
C_{ij}\sqrt{\lambda_i\lambda_j}I_1(\lambda_i,\lambda_j)\nonumber\\
&&+\ \frac{1}{2}(\lambda_i-\lambda_j)C^\ast_{ij}I_3(\lambda_i,\lambda_j)
\Bigg]\nonumber\\
&&+\ \frac{1}{2} \Delta^A_{ij}(\lambda_i-\lambda_j)C^\ast_{ij}
\Big[\ I_2(\lambda_i,\lambda_j)-I_1(\lambda_i,\lambda_j)\ \Big] , \\[0.5cm]
{\cal T}^{l_1l_2}_b\ &=&\ \frac{1}{2}\Delta^S_{ij}\ \Bigg[
\Big(\mbox{C}_{UV}-\frac{1}{2}\Big) \Big( \lambda_i\delta_{ij}-
C_{ij}\sqrt{\lambda_i\lambda_j}-\frac{1}{2}(\lambda_i+\lambda_j)C^\ast_{ij}\Big)
\ -\ (\lambda_i+\lambda_j)(\delta_{ij}\nonumber\\
&&-\ C^\ast_{ij})L_2(\lambda_i,\lambda_j)
\ +\ C_{ij}\sqrt{\lambda_i\lambda_j}\Big(\ -2L_2(\lambda_i,\lambda_j)\ +\
\frac{1}{2}(\lambda_j-\lambda_i)I_3(\lambda_i,\lambda_j)\ \Big) \Bigg]
\nonumber\\
&&+\ \frac{1}{2}\Delta_{ij}^A(\lambda_i-\lambda_j)\Big( \  \frac{1}{2}
(\mbox{C}_{UV}-\frac{1}{2})C^\ast_{ij}\ -\ C_{ij}\sqrt{\lambda_i\lambda_j}
I_2(\lambda_i,\lambda_j)\nonumber\\
&&+\ C_{ij}^\ast L_2(\lambda_i,\lambda_j)\ \Big) ,
\\[0.5cm]
{\cal T}^{ff}_c\ &=&\  -2\delta^S(2T^f_z)\Big(\lambda_j C_{ij}(
\delta_{ij}- C^\ast_{ij})\ -\ \sqrt{\lambda_i\lambda_j} \mbox{Re}C_{ij}^2
\Big)\Big( \ - \frac{1}{2}\mbox{C}_{UV}\nonumber\\
&& +\ L_1(\lambda_i,\lambda_j)\ \Big) ,\\[0.5cm]
{\cal T}^{ff}_d\ &=&\ -i\delta^A\mbox{Im}C_{ij}^2(c_\theta-s_\theta t_\beta)
(\lambda_i-\lambda_j)\sqrt{\lambda_i\lambda_j}\lambda_H^{-1}
\Big( \ \mbox{C}_{UV}\ -\ L_0(\lambda_i,\lambda_j)\ \Big),\\[0.5cm]
{\cal T}^{ff}_e\ &=&\  i\delta^A\mbox{Im}C_{ij}^2(s_\theta+c_\theta t_\beta)
(\lambda_i-\lambda_j)\sqrt{\lambda_i\lambda_j}\lambda_S^{-1}
\Big( \  \mbox{C}_{UV}\ -\ L_0(\lambda_i,\lambda_j)\ \Big) ,
\eea 
where
\bea
{\lambda}_i &=& \frac{m_{n_i}^2}{M^2_W} \ , \qquad
\lambda_H=\frac{M^2_H}{M^2_W},
\qquad \lambda_S=\frac{M^2_S}{M^2_W},\\
{\mbox{C}}_{UV} & = & \frac{1}{\varepsilon} - {\gamma}_E + \ln 4\pi -
\ln \frac{M^2_W}{\mu^2}, \\
\Delta^A_{ij} & = & -\ \frac{g_W{\alpha}_W}{16\pi} \tan\beta
\ B_{l_1i}^\ast B_{l_2j}
\ \ {\bar{\mbox{u}}}_{l_2} \left[ \frac{m_{l_1}}{M_W}(1+\gamma_5)
\ +\ \frac{m_{l_2}}{M_W}(1-\gamma_5) \right]
{\mbox{u}}_{l_1}, \nonumber\\
\Delta^S_{ij} & = & -\ \frac{g_W{\alpha}_W}{16\pi} \tan\beta
\ B_{l_1i}^\ast B_{l_2j}
\ \ {\bar{\mbox{u}}}_{l_2} \left[ \frac{m_{l_1}}{M_W}(1+\gamma_5)
\ -\ \frac{m_{l_2}}{M_W}(1-\gamma_5) \right]{\mbox{u}}_{l_1}\ , \\[0.4cm]
\delta^A & =& -\ \frac{g_W\alpha_W}{16\pi}\ \tan\beta\
\bar{\mbox{u}}_f\mbox{u}_f\ ,\nonumber\\
\delta^S & =& -\ \frac{g_W\alpha_W}{16\pi}\ \tan\beta\
\bar{\mbox{u}}_f\gamma_5\mbox{u}_f\ .
\eea  
The analytical expressions for the one-loop functions
$I_1$, $I_2$, $I_3$, $L_0$, $L_1$ and $L_2$ are given in Appendix~A.
In Eq.~(20) $T^f_z$ stands for the third component of the weak isospin and
takes the values: $T^u_z=1/2$, $T^{l,d}_z=-1/2$. The $UV$~divergences in the
amplitudes~(19)--(22) vanish identically due to the following
equalities~\cite{ZPC,KPS}:
\bea
\sum\limits_{i=1}^{2n_G} B_{li}C_{ij} & = & \ B_{lj}\ , \\
\sum\limits_{k=1}^{2n_G} C_{ik}C_{jk}^\ast & = & C_{ij}\ , \\
\sum\limits_{i=1}^{2n_G} m_{n_i} B_{li}C^{\ast}_{ij} & = & \ 0\ , \\
\sum\limits_{k=1}^{2n_G} m_{n_k}C_{ik}C_{jk} & = & \ 0\ .
\eea 
Note also that the amplitudes ${\cal T}_d^{ff}$, ${\cal T}^{ff}_e$ induce
a scalar piece in the $J^0-f-f$ coupling. This scalar part, however, is
suppressed for astrophysical reasons by a factor of 10 at least as compared to
the pseudoscalar part of the coupling~\cite{GMP}.

In order to pin down numerical predictions, we first consider the
conservative case of a model with one generation or
equivalently a three-generation
model without interfamily mixings. Then,
the coupling $J^0-e-e$, $\mbox{g}_{Jee}$, defined by the relation
\beq
{\cal T}^{ee}\ \ =\ \ \mbox{g}_{Jee}\ \bar{e}i\gamma_5e,
\eeq 
takes the simple form
\beq
\mbox{g}_{Jee}\ \simeq\ \frac{g_W\alpha_W}{16\pi}t_\beta\frac{m_e}{M_W}\
\left[ (s^{\nu_e}_L)^2 \frac{\lambda_{N_e}^2}{1-\lambda_{N_e}}
\left(1\ +\ \frac{\ln\lambda_{N_e}}{1-\lambda_{N_e}}\right)\ +\
\frac{1}{2}\sum\limits_{e,\mu,\tau} (s^{\nu_l}_L)^2 \lambda_{N_l} \right].
\eeq 
The mixings $(s^{\nu_l}_L)^2$ are defined by
\beq
(s^{\nu_l}_L)^2\ =\ \sum\limits_{i=n_G+1}^{2n_G}\ |B_{li}|^2\ .
\eeq 
In this scenario $(s^{\nu_l}_L)^2=m^2_{D_{ll}}/m^2_{N_l}$ and $\lambda_{N_l}
\gg 1$, since the heavy Majorana neutrinos $N_l$ have to satisfy Eq.~(1) and
$m_{D_{ll}} \simeq m_{l_i}$ or $m_{u_i}$. On the other hand, astrophysical
constraints arising from helium ignition in red giants or the observational
evidence of white dwarf cooling rates are given by the bound~\cite{GR}
\beq
\mbox{g}_{Jee}\ \  \leq\ \  (9.-1.4)\ 10^{-13}\ ,
\eeq 
However, the range $3.\ 10^{-13} \leq \mbox{g}_{Jee} \leq 6.\ 10^{-7}$
is excluded from the helium ignition argument mentioned above, if the radius
of giant core or dwarf is bigger than the mean free path of the pseudoscalars
that these particles require to freely escape from them~\cite{GR}.
It is now obvious that the most stringent constraint on $\tan\beta$
arises from the heaviest family.
Thus, for~$m_D\simeq m_\tau$ one obtains that
\beq
\mbox{g}_{Jee} \ \simeq \ \frac{g_W \alpha_W}{32} \tan\beta\ \frac{m_e}{M_W}
\frac{m^2_\tau}{M^2_W}\ ,
\eeq 
yielding because of Eq.~(34)
\beq
\tan\beta\ \ \leq \ \ 0.4\ .
\eeq 
Of course, if $m_D\simeq m_t/k \simeq 10$~GeV, one finds a much stronger
bound, i.e.
\beq
\tan\beta\ \ \leq \ \ 10^{-2}\ .
\eeq 
Note also that such low-energy realizations make unlikely the invisible decay
of massive Higgses into Majoron pairs~\cite{JR,JV}.

The afore-mentioned hierarchical scheme, however, is in general not valid
if one introduces intergenerational mixings in the singlet Majoron model.
This situation seems to be a natural possibilty that can be realized by $GUT$
models, since $m_D$ and $M_U$ matrices  may get related in such high-energy
scenarios (i.e. $m_D(M_X)=M_U(M_X)$ with $M_X$ indicating the grand
unification scale). In addition, it has explicitly been
demonstrated in~\cite{BW,ZPC}
that the scale of $m_M$ can be ${\cal O}(100)$~GeV without contradicting
experimental bounds on neutrino masses.
For instance, democratic-type mass matrices for the form of $m_D$~\cite{DEM}
can lead to patterns with such a low scale for $m_M$. Then, the mixings
$(s^{\nu_l}_L)^2$ can be treated as purely phenomenological parameters,
since Eq.~(33)  should now read
\beq
(s^{\nu_l}_L)^2\ \ \simeq \ \ m_D\ \frac{1}{m^2_M}\ m_D^\dagger
\eeq 
and cannot therefore be related with the light-neutrino mass matrix of Eq.~(1).
The mixing angles $(s^{\nu_l}_L)^2$ can generally be constrained by a
global analysis of a great number of low-energy experiments and $LEP$
data~\cite{LL}.
In this scenario one makes the remerkable observation that $\mbox{g}_{Jee}$ can
severely be supressed for a certain choice of the mass parameters
$\lambda_{N_l}$ and mixings $(s^{\nu_l}_L)^2$. For example, if all heavy
neutrino masses $m_{N_l}$ are approximately equal and
$\lambda_{N_l} \gg 1$, then the choice
\beq
(s^{\nu_e}_L)^2\ \  \simeq \ \  (s^{\nu_\tau}_L)^2
\eeq 
leads to $\mbox{g}_{Jee} = 0$. However, even if the Majoron couplings to
electrons vanish, the corresponding coupling to nucleons ${\cal N}$,
$\mbox{g}_{J{\cal N\cal N}}$, is not zero anymore. The reason is that the
destructive first term in the bracket of Eq.~(32) does not exist anymore and
such a fine-tuning is thus not possible. Since $\mbox{g}_{J{\cal N\cal N}}/
\mbox{g}_{Jee}\simeq m_{\cal N}/m_e \simeq 2.\ 10^3$, one may
derive useful constraints from the consideration of cooling rates of neutron
stars due to the energy loss mechanism by Majoron emission.
In Fig.~(3) we present
exclusion plots of the parameters $\tan\beta$ versus $m_N$ for three
different values of $(s^{\nu_l}_L)^2$ by considering
that~\cite{IW}
\beq
\mbox{g}_{J{\cal N\cal N}}\ \ \stackrel{\displaystyle <}{\sim} \ \ 10^{-9}.
\eeq 
For a discussion of additional uncertaintities on the upper bound of the
coupling $\mbox{g}_{J\cal N\cal N}$
that can arise from various reasons like the so-called
"Turner's window"~\cite{TUR} etc., we refer the reader to~\cite{GGR}.
Ultimately, we must notice that the astrophysical
bounds arising from the Majoron coupling to two photons, $C_{J\gamma\gamma}$,
should be weaker than that coming from $\mbox{g}_{J{\cal N\cal N}}$, since
$C_{J\gamma\gamma}$ can only be generated at two-loop electroweak
order.

In the following we will focus our attention on bounds resulting solely from
terrestrial experiments~\cite{TER1,TER2} by analyzing lepton-flavor violating
decays, i.e.~$l_1\to J^0 l_2$ with $l_1\neq l_2$. To the leading order
of the heavy neutrino limit one finds from Eqs.~(18) and (19) that
\beq
\mbox{BR}(l_1^- \to J^0 l_2^-)\ \simeq\ \frac{3\alpha_W}{8\pi}
\tan^2\beta\ |B^\ast_{l_1N} B_{l_2N} |^2 \lambda^2_N\ \frac{M^2_W}{m^2_{l_1}}.
\eeq 
The experimental information we have for the above lepton-violating decays
are the following upper bounds:
\bea
\mbox{BR}(\mu \to J^0 e)\ &\leq&\ 2.6\ 10^{-6}\ \mbox{\cite{TER1}}, \nonumber\\
\mbox{BR}(\tau \to J^0 e)\ &\leq&\ 7.1\ 10^{-3}\ \mbox{\cite{TER2}},
\nonumber\\
\mbox{BR}(\tau \to J^0 \mu)\ &\leq&\ 2.3\ 10^{-3}\ \mbox{\cite{TER2}}.
\eea 
In order to quantitatively estimate the magnitude of the lepton-violating
effects that could be
constrained by the branching ratios stated in~(42), we use the upper bound
of the quantity
\beq
|B^\ast_{l_1N} B_{l_2N} |\ \leq\ (s^{\nu_l}_L)^2\ =\ \mbox{max}\Big(
(s^{\nu_{l_1}}_L)^2,\ (s^{\nu_{l_2}}_L)^2\Big).
\eeq 
The exclusion plots implied by these experiments are presented in Fig.~(4)
for the three different decay channels. For comparison, we have taken the
astrophysical bound coming from Eq.~(40) into account in Fig.~(4), from
which one easily concludes that experimental searches for the decay
$\mu \to J^0e$ may not be
excluded by astrophysical constraints and can hence be sensitive to new
physics beyond the $SM$.

In conclusion, astrophysical considerations may lead to useful constraints
on the parameters of singlet Majoron models with intergenerational mixings.
It has been demonstrated that three-generation Majoron models can indeed be
constrained if these models are assumed to be embedded in $GUT$ scenarios.
Possibilities of how to evade from some of the astrophysical constraints have
also been discussed. For example, $\mbox{g}_{Jee}$ vanishes for a specific
choice of parameters. Furthermore, terrestrial experiments give
independently severe restrictions on the lepton-violating mixings and heavy
neutrino masses. Aside from rather involved $R$-parity broken
models~\cite{RRV},
this minimal extension of the $SM$, the singlet Majoron model, may also
naturally account for possible lepton-flavor violating signals in precision
experiments. We emphasize again the fact that measurements of the $TRIUMF$
collaboration~\cite{TER1} for exotic decay modes, like $\mu \to J^0 e$,
lie in area which may not be excluded by astrophysics and have substancial
chances to establish new physics beyond the $SM$.
Finally, due to the $CP$-odd interactions
that Majorona neutrinos introduce in such models (see e.g.~Eqs.~(9), (14) and
(15)), one may be motivated to discuss their phenomenological impact
of possible $CP$-violating effects in the decays of the Higgs particle
$H^0$ into top, W or Z pairs~\cite{IKP}.\\[1.cm]
{\bf Acknowledgements.} I wish to thank E.A.~Paschos, W.~Buchm\"uller,
M.~Nowakowski, B.~Kniehl, A.~Ilakovac and S.~Rindani for helpful discussions.
This work has been supported by a grant from the Postdoctoral Graduate College
of Mainz.

\newpage

\setcounter{section}{1}
\setcounter{equation}{0}
\def\theequation{\Alph{section}\arabic{equation}}

\begin{appendix}
\section{The loop integrals}
\indent

We first define the useful functions
$B_1(\lambda_i,\lambda_j)$ and $B_2(\lambda_i,\lambda_j)$ as:
\bea
B_1(\lambda_i,\lambda_j) &\  = \ & \lambda_i(1-x) + \lambda_j\, x\ , \\
B_2(\lambda_i,\lambda_j) &\ = \ & 1-y+y[\lambda_i(1-x)+\lambda_j\, x] \ ,
\eea 
where $x$ and $y$ are Feynman parameters.
The loop integrals $L_0$, $L_1$, $L_2$, $I_1$, $I_2$ and $I_3$
are then given by
\bea
L_0(\lambda_i,\lambda_j)\ &=&\ \int dx\, \ln B_1(\lambda_i,\lambda_j)\
\nonumber\\
&=&\ -1\ +\ \frac{1}{2}\ln\lambda_i\lambda_j\ -\ \frac{\lambda_i
+\lambda_j}{2(\lambda_i-\lambda_j)}\ln\frac{\lambda_j}{\lambda_i}\, ,\\
L_1(\lambda_i,\lambda_j)\ &=&\ \int dx\, x\ln B_1(\lambda_i,\lambda_j)\
\nonumber\\
&=&\ -\frac{1}{4}\ +\ \frac{\lambda_i^2}{2(\lambda_i-\lambda_j)^2}
\ln\frac{\lambda_i}{\lambda_j}\ +\ \frac{1}{2}\ln\lambda_j\ -\
\frac{\lambda_i}{2(\lambda_i-\lambda_j)}\, , \\
L_2(\lambda_i,\lambda_j)\ &=&\ \int dx dy \ y\ln B_2(\lambda_i,\lambda_j)\
\nonumber\\
&=&\ \frac{3}{4(\lambda_i-\lambda_j)}\ \left[ \frac{\lambda_j}{1-
\lambda_i}\ -\ \frac{\lambda_i}{1-\lambda_j} \right]\ -\
\frac{3\lambda_i\lambda_j}{4(1-\lambda_i)(1-\lambda_j)}\ \nonumber\\
&&+\ \frac{1}{2(\lambda_i-\lambda_j)}\left[ \frac{\lambda_j^2\ln\lambda_j}
{1-\lambda_j}\ -\ \frac{\lambda_i^2\ln\lambda_i}{1-\lambda_i} \right]\, ,\\
I_1(\lambda_i,\lambda_j)\ &=&\ \int \frac{dxdy\, y}{B_1(\lambda_i,
\lambda_j)}
\nonumber\\
&=&\ \frac{\lambda_i\lambda_j\ln(\lambda_i/\lambda_j)+\lambda_j\ln\lambda_j
-\lambda_i\ln\lambda_i}{(1-\lambda_i)(1-\lambda_j)(\lambda_i-\lambda_j)}\,
,\\
I_2(\lambda_i,\lambda_j)\ &=&\ \int\frac{dxdy\ y^2}{B_2(\lambda_i,\lambda_j)}
\nonumber\\
&=&\ -\frac{1}{2(1-\lambda_i)(1-\lambda_j)}\ +\ \frac{1}{2(
\lambda_i-\lambda_j)}\left[ \ln\frac{\lambda_i}{\lambda_j}\ -\
\frac{\ln\lambda_i}{(1-\lambda_i)^2}\right. \nonumber\\
&&\left. +\ \frac{\ln\lambda_j}{(1-\lambda_j)^2}
\right] \,  ,\\
I_3(\lambda_i,\lambda_j)\ &=&\ \int\frac{dxdy\ y^2(1-2x)}{B_2(\lambda_i,
\lambda_j)} \nonumber\\
&=&\ -\frac{1}{2(1-\lambda_i)(1-\lambda_j)}\left[ \frac{\ln\lambda_i}{
1-\lambda_i}\ -\ \frac{\ln\lambda_j}{1-\lambda_j} \right]\ -\
\frac{1}{2(\lambda_i-\lambda_j)}\left[ \frac{\lambda_j}{1-\lambda_j}
\right. \nonumber\\
&&\left.  +\
\frac{\lambda_i}{1-\lambda_i} \right]
\ -\frac{\ln(\lambda_j/\lambda_i)}{2(\lambda_i-\lambda_j)^2}
\left[ \frac{\lambda_j^2}{1-\lambda_j}\ +\ \frac{\lambda_i^2}{1-\lambda_i}
\right]\, .
\eea 
The integration interval of the variables $x$ and $y$ is $[0,1]$.

\end{appendix}

\newpage

\centerline{\bf\Large Figure Captions }
\vspace{1cm}
\newcounter{fig}
\begin{list}{\bf\rm Fig. \arabic{fig}: }{\usecounter{fig}
\labelwidth1.6cm \leftmargin2.5cm \labelsep0.4cm \itemsep0ex plus0.2ex }

\item Feynman graphs responsible for the coupling of Majorons to fermions,
$J^0-f_1-f_2$.

\item $CP$-odd graphs giving rise to a scalar part in the coupling
$J^0-f-f$.

\item Exclusion plots from astrophysical requirements.
We have considered the values: $(s^{\nu_l}_L)^2=5.\ 10^{-2}$ (solid line),
$(s^{\nu_l}_L)^2=10^{-2}$ (dashed line), $(s^{\nu_l}_L)^2=10^{-3}$
(dot-dashed line). The area lying above of the curves is excluded by the
restriction $\mbox{g}_{J\cal N\cal N} <10^{-9}$. In addition, we assume
that all heavy neutrino masses are approximately equal with $m_N$.

\item Exclusion plots originating from the decays:
$\mu \to J^0 e$ (solid line), $\tau \to J^0 e$ (dashed line),
$\tau \to J^0 \mu$ (dot-dashed line). For comparison, we have considered
the astrophysical bound $\mbox{g}_{J\cal N\cal N} \leq 10^{-9}$ (dotted
line). The areas lying above of the curves are excluded by the afore-mentioned
conditions.

\end{list}

\end{document}